\documentclass[conference]{IEEEtran}
\IEEEoverridecommandlockouts

\usepackage{multirow}
\usepackage{makecell}
\usepackage{siunitx}
\usepackage[table,dvipsnames]{xcolor}
\usepackage{tikz}
\usepackage{pgfplots}
\pgfplotsset{compat=1.17}
\usepgfplotslibrary{groupplots}
\usetikzlibrary{patterns}
\usepackage{amsmath,amssymb}
\usepackage{algorithm}
\usepackage{algpseudocode}
\DeclareMathOperator*{\argmax}{arg\,max}

\usepackage{cite}
\usepackage{url}
\usepackage{amsmath,amssymb,amsfonts}
\usepackage{graphicx}
\usepackage{textcomp}
\usepackage{xcolor}
\usepackage{algorithm}
\usepackage{algpseudocode}
\usepackage{booktabs,graphicx}
\usepackage{tabularx}
\def\BibTeX{{\rm B\kern-.05em{\sc i\kern-.025em b}\kern-.08em
    T\kern-.1667em\lower.7ex\hbox{E}\kern-.125emX}}

\DeclareUnicodeCharacter{2068}{}
\providecommand{\PH}[1]{#1}

\makeatletter
\newcommand{\linebreakand}{%
  \end{@IEEEauthorhalign}%
  \hfill\mbox{}\par
  \mbox{}\hfill\begin{@IEEEauthorhalign}%
}
\makeatother

\begin{document}
\bstctlcite{IEEEcontrol}
\title{RegionAlign: Manipulating Parameter Layouts for Localized Fault Injection on LLMs and VLMs \ \
}

\author{\IEEEauthorblockN{Xiaomeng Wang}
\IEEEauthorblockA{
\textit{Radboud University}\\
Nijmegen, Netherlands \\
xiaomeng.wang@ru.nl}
\and
\IEEEauthorblockN{⁨Péter Horváth}
\IEEEauthorblockA{
\textit{Radboud University}\\
Nijmegen, Netherlands \\
peter.horvath@ru.nl
}
\and
\IEEEauthorblockN{Zhengyu Zhao}
\IEEEauthorblockA{
\textit{Xi'an Jiaotong University}\\
Xi'an China \\
zhengyu.zhao@xjtu.edu.cn}
\linebreakand
\IEEEauthorblockN{Martha Larson}
\IEEEauthorblockA{
\textit{Radboud University}\\
Nijmegen, Netherlands \\
martha.larson@ru.nl}
\and
\IEEEauthorblockN{Zhuoran Liu}
\IEEEauthorblockA{
\textit{University of Amsterdam}\\
Amsterdam, Netherlands \\
z.liu@uva.nl}
\and
\IEEEauthorblockN{Lejla Batina}
\IEEEauthorblockA{
\textit{Radboud University}\\
Nijmegen, Netherlands \\
lejla@cs.ru.nl}
}

\maketitle

\begin{abstract}

Physical fault injection can flip stored weight bits, degrading LLMs and VLMs. Existing work largely overlooks how supply-chain manipulation affects fault vulnerability. We introduce RegionAlign, which allows a malicious provider to manipulate only the parameter layout before deployment, using bit-sensitivity search and channel permutations to concentrate vulnerable bits in one memory region while preserving weight values and clean performance. 
Under matched budgets, software experiments show greater degradation from search-selected in-region flips with RegionAlign than with original or random layouts.
EMFI feasibility study confirms weight corruption.
RegionAlign reveals a stealthy supply-chain threat that layout manipulation weakens physical fault resilience.

\end{abstract}


\section{Introduction}
\label{sec:introduction}
Deploying third-party Large Language Models (LLMs) and Vision-Language Models (VLMs) introduces supply-chain security risks: providers control the models they distribute, and users who check only task performance may overlook malicious modifications that preserve clean behavior~\cite{gu2017badnets, kurita-etal-2020-weight}. 
A separate threat arises in hardware: fault injection corrupts the stored weights of the deployed model to degrade model performance, and bit-flip attacks (BFAs) are its main realization, flipping a small number of weight bits~\cite{rakin2019bfa,das2025attentionbreaker,yao2020deephammer}. 
Bit flips in the stored weights have been studied under different threat models: untargeted inference-time degradation after model deployment, and backdoors implanted by bit flips, which are activated later either by an input trigger~\cite{li2025oneflip} or by the fault itself~\cite{hommes2026triggering}.

These two threats, model supply-chain attacks and physical fault-injection attacks after deployment, have so far been studied in isolation: supply-chain attacks rarely consider the physical constraints of fault injection, while physical attacks typically assume a fixed parameter layout or rely on an input trigger~\cite{yao2020deephammer,li2025oneflip,roy2026robbin,lin2025gpuhammer}. 
This separation leaves a practical aspect unaddressed: vulnerable bits identified across a model can be widely scattered in memory~\cite{das2025attentionbreaker,khalil2026flipllm,yao2020deephammer,lin2025gpuhammer}, whereas a localized physical fault injection can affect only a limited region. 
Consequently, even attacks requiring only a few bit flips may demand multiple injection locations~\cite{yao2020deephammer,roy2026robbin}, making physical realization far harder than a single localized injection.

\begin{figure}
    \centering
    \includegraphics[width=\linewidth]{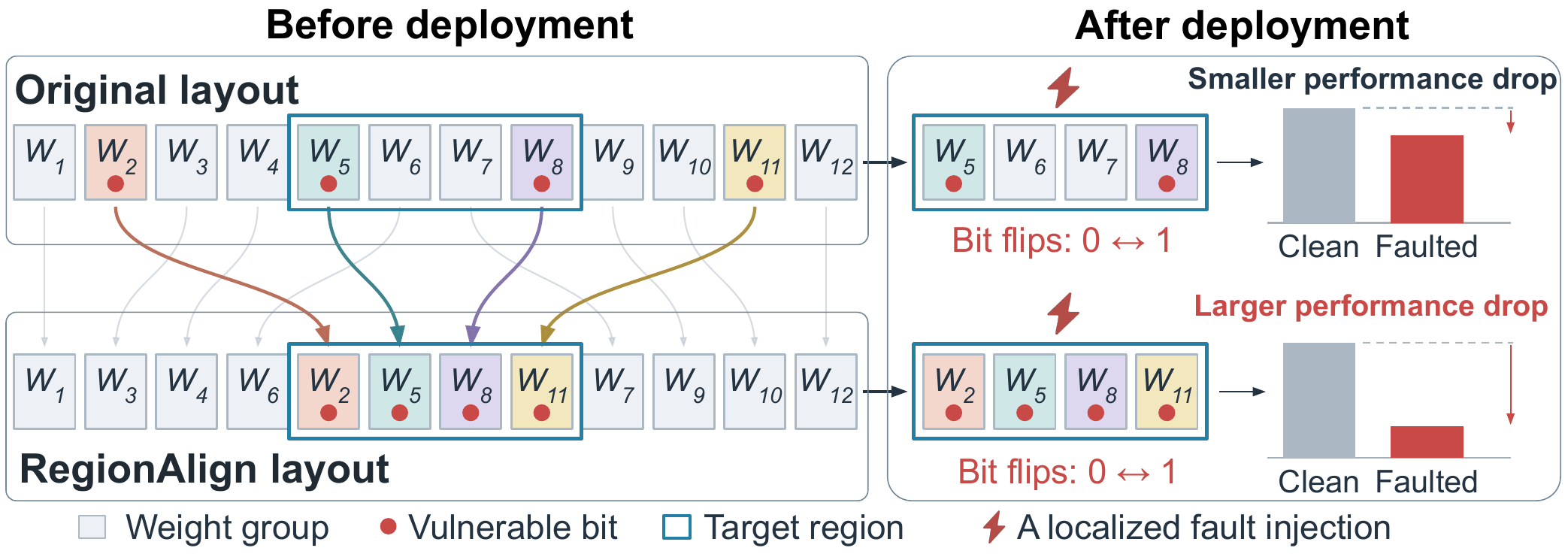}
    \caption{RegionAlign attack overview. Before deployment, a malicious provider uses channel permutations to concentrate vulnerable weight bits in a target region while keeping clean behavior nearly unchanged. In-region bit search may select different bits across layouts. After deployment, one localized fault injection affects the same region in each layout. The RegionAlign layout causes greater task-performance degradation. }
    \label{fig:overview}
\end{figure}
In this paper, we identify the parameter layout as a critical link between supply-chain manipulation and physical fault injection. 
The parameter layout is the mapping from logical weights to their storage positions, independent of the weight values themselves.
We focus on bit-flip attacks. 
A malicious provider can rearrange LLM and VLM weights before the model's publication to concentrate vulnerable bits within a physically clustered region while preserving clean (fault-free) model behavior. 
In the original layout, vulnerable bits are mostly scattered across memory. 
In the rearranged layout, vulnerable bits can be concentrated within a fixed target region through channel permutations that leave clean model behavior nearly unchanged. 
After deployment, an attacker with the required physical access performs one physical fault injection in the single target region to cause the degradation. 
Figure~\ref{fig:overview} illustrates the attack.

To find vulnerable parameter layouts, we introduce RegionAlign, an attack that jointly searches admissible layouts and damaging bit flips within a fixed target region.
For each candidate layout, we rank flips by estimated effects and select them by measured loss increases.
We adapt channel permutations used for structured sparsity~\cite{pool2021permutations} to LLM/VLM attention and quantization constraints, preserving weight values and model function in exact arithmetic.
Preparation requires no retraining and leaves the inference pipeline unchanged, and each layout must pass independent checks of clean task performance.
Together, these steps yield parameter layouts that preserve clean model performance while making selected in-region bit flips more damaging than under the original layout.

We first evaluate RegionAlign through software fault injection, applying search-selected bit flips directly to the INT8 weights of LLMs and VLMs. 
Under matched target regions, bit-flip budgets, and search budgets, we compare the original layout, the attacker-chosen RegionAlign layout, and a randomly sampled admissible layout outside the attacker's control.
All three layouts retain the same weight values before corruption and have nearly identical clean task performance.
These software-injected flips can cause greater degradation under RegionAlign than under either baseline.
For example, a single in-region flip raises StableLM2-1.6B's WikiText-2 perplexity from 11.53 to 68.29, while both baselines stay near 11.5.
These results show that functional equivalence of the model does not guarantee resilience to localized weight faults, and that clean performance evaluation alone can miss this vulnerability.
We then examine the physical realizability. Electromagnetic fault injection (EMFI) on an NVIDIA Jetson AGX Thor corrupts INT8 weights in DRAM in direction-agnostic multi-bit bursts, consistent with our threat model. Yet none of $2500$ attempts reached the pre-registered target tensor, highlighting the difficulty of region targeting. This supports our claim that the region constrains the attacker and that the attack's leverage comes from the layout within it rather than unconstrained physical faults.

To summarize, this work makes the following contributions:

\begin{itemize}
    \item We uncover a physically-aware supply-chain attack vector: before deployment, channel permutations concentrate vulnerable bits within a fixed memory region, so that a single localized fault injection compromises the model while leaving clean behavior nearly unchanged.
    \item We propose RegionAlign, which combines sensitivity-guided layout search with channel permutations to recognize and then concentrate vulnerable weight bits in a target region without retraining or changing weight values.
    \item Across INT8-quantized LLMs and VLMs, experimental results demonstrate that selected in-region flips cause greater performance degradation with RegionAlign than with the original or a random admissible layout.
    \item A hardware feasibility study with EMFI on an NVIDIA Jetson AGX Thor showing that weight-bit faults in DRAM-resident quantized models are physically realizable and direction-agnostic, while un-steered injection reaches a chosen region with near-zero probability, corroborating our region-constrained threat model.
\end{itemize}

\section{Background and Related Work}
\noindent \textbf{LLMs and VLMs.}
LLMs generate text using Transformer decoders; VLMs incorporate image features~\cite{touvron2023llama2,liu2024llava15}.
Their attention and feed-forward networks (FFNs) use linear projections between feature dimensions (channels)~\cite{vaswani2017attention}.
Weights are parameter values indexed by logical coordinates
(layer, tensor, row, column); activations are input-dependent features.
Our checkpoints store INT8 weight codes, scales, and other metadata.
Parameter layout maps logical weights to storage positions. Deployment determines their physical addresses.

\noindent \textbf{Bit-flip attacks.}
Existing attacks identify damaging flips in fixed layout ~\cite{das2025attentionbreaker,khalil2026flipllm}.
SBFA shows single-bit LLM degradation while keeping the corrupted weight within its layer's benign range~\cite{guo2025sbfa}.
DeepHammer incorporates device-specific flip capability and page placement~\cite{yao2020deephammer}; GPUHammer extends Rowhammer to GPU memory~\cite{lin2025gpuhammer}.
Beyond attacking fixed models, training-assisted methods prepare models to require fewer damaging flips~\cite{dong2023onebit}.
Input-triggered backdoors are installed through weight-bit flips~\cite{li2025oneflip,roy2026robbin}, with ROBBIN accounting for collateral flips. Fault-induced perturbations can instead activate feature-map
backdoors~\cite{hommes2026triggering}.

\noindent  \textbf{Comparison with prior work.}
RegionAlign extends fixed-layout bit search with function-preserving channel permutations~\cite{pool2021permutations} to concentrate vulnerable bits in a fixed target region.
Unlike training-assisted preparation, it requires neither retraining nor changes to represented weight values.
Its objective is untargeted degradation from in-region weight corruption, without an input trigger.

\section{Threat Model}
\label{sec:threat_model}
\noindent  \textbf{Setting.}
We consider a quantized model supplied by a malicious model provider, attacked in two stages. 
Before deployment, the provider prepares the checkpoint (the stored parameters and metadata), so that its clean performance (its performance with no fault) is unchanged, yet localized physical fault injection at one location, mapped to a fixed target region, becomes more damaging than on the original checkpoint.
After deployment, the attacker or a colluding party with injection access performs one fault injection at that single location to activate a chosen downstream objective, e.g., performance degradation. 
It's not easy for an attacker to precisely control which bits a fault injection flips.

\noindent \textbf{Adversarial goal.}
The bit flip induced by the preparation is an outcome, rather than the attack objective itself. 
Depending on the adversary’s goal, it may be used to degrade model performance, bypass safety safeguards (i.e., jailbreak), or crash the workload. 
In this work, we focus on performance degradation.
The attacker aims to increase the probability of such degradation from a single injection at the fixed location, relative to the original checkpoint under the same injection conditions.

\noindent\textbf{Before deployment: checkpoint preparation.}
The attacker has access to the model architecture, weight values, quantization format, and parameter layout (where the weight values are stored), but the attacker cannot change any parameter values. 
The only permitted change is the parameter layout, applied by channel permutations that keep the weight values and the model function identical in exact arithmetic (See Section~\ref{sec:method}).
Retraining, requantization, arbitrary weight-value changes, and inference-pipeline modifications are excluded.


\noindent \textbf{During deployment: Target region and deployment mapping.}
The region affected by a localized fault injection depends on the hardware model and injection conditions and may vary across devices. 
Following common practice in fault-injection research~\cite{van2020fill}, we assume that the attacker has access to a copy of the target hardware for prior fault characterization, identifying an effective injection location, the affected region, and the corresponding fault model. 
We assume that the characterized fault behavior transfers to the victim device under the same conditions.
As in prior physical memory attacks~\cite{yao2020deephammer,lin2025gpuhammer}, the attacker must also reverse-engineer the mapping between weight memory addresses and physical storage locations to determine which encoded weight bits occupy the affected region. 
We assume this mapping is known and stable and that deployment preserves the prepared weight representation and its mapped placement. 
Access to a model checkpoint alone does not establish or control this physical placement.

\noindent  \textbf{After deployment: fault injection preparation.}
The attacker has the access needed to inject a fault at the
hardware location associated with the target region.
An injection attempt contains one localized injection with multiple tries.
For an EMFI realization, a single pulse under the specified conditions. 
The corruption affects bits of stored quantized weight within the target region
and persists over the inference window. 
On platforms using AES-XTS memory encryption, faults that alter ciphertext and survive hardware error handling can produce multi-bit changes in the corresponding plaintext weight blocks.
This provides one hardware motivation for not assuming precise control over individual bit locations or flip directions.

\section{RegionAlign}
\label{sec:method}
RegionAlign in Algorithm~\ref{algorithm} searches for a parameter layout that makes bit flips inside a
fixed target region $R$ more damaging. Given a target region $R$, RegionAlign returns one permuted layout and a bit-flip set for each budget $K$.

\subsection{Channel Permutations Preserving the Function}
A channel permutation consistently reorders channels across connected operators, which changes the layout but neither the weights nor the computed function~\cite{pool2021permutations}.
It is admissible if it preserves the model function:
\begin{itemize}
\item \textbf{Feed-forward.} In gated blocks~\cite{shazeer2020glu}, the output
rows of both input projections and the input columns of the output projection
are reordered together.
\item \textbf{Attention.} Head permutations move complete key--value sharing
groups~\cite{ainslie2023gqa} and update the query, key, value, and output
projections together. Value coordinates may be reordered identically within a
sharing group; query and key coordinates within a head are fixed.
\item \textbf{Residual stream.} Embeddings, projections, and normalization
parameters are reordered together.
\item \textbf{Biases and shared parameters.} A projection bias follows the
output-axis order of its weight. A shared parameter is permuted once per axis.
\item \textbf{Quantization.} Weight codes move with their scales, column
permutations keep quantization groups intact, and nothing is requantized.
\end{itemize}
We implement these parameter permutations. Other per-channel
parameter permutations require equivalence checks.

\subsection{Candidate Layouts and Flip Scoring}
Search runs in the original layout: a flip applied before an admissible permutation equals its mapped flip applied after it, which deployment validation checks separately.

\paragraph{Flip scores}
Search first measures the clean model's attack loss, the gradient of the target weight tensor, and its input activations. For a flip with decoded weight change $\Delta w$, weight gradient $g$, and maximum absolute input activation $a$ of its column,
\[
s_{\mathrm{mag}} = |\Delta w|\,a,
\qquad
s_{\mathrm{fo}} = g\,\Delta w
\]
give an activation-weighted magnitude and a first-order loss estimate. A flip is allowed only if the resulting weight is finite and within the row's in-range limit, i.e., the maximum absolute weight of that row in the clean model.
For each weight, we keep the best allowed bit per score.
Scores propose flips; measured attack loss changes decide acceptance.

\begin{algorithm}[t]
\caption{RegionAlign}
\begin{algorithmic}[1]
\Require model, region $R$, attack set, computational budget $B$, stages
$(c,\kappa,\phi)$ ending at $\kappa=K_{\max}$
\Ensure permutation $\pi^\star$, flip sets $\{F_k\}_{k=1}^{K_{\max}}$
\State Measure clean loss, gradient, and activations
\State $V \gets \{G_{\mathrm{orig}}\} \cup \{\text{top reachable rows}\}$,
each with its queue and surrogate
\For{each stage $(c,\kappa,\phi)$}
    \State Keep the top $c$ of $V$ by surrogate (first stage) or
    $\operatorname{AUC}$
    \State $\beta \gets \lfloor \phi\,(B-\mathrm{spent})/|V| \rfloor$
    \For{$G \in V$}
        \State \textsc{LazyGrow}$(G,\kappa,\beta)$
        \Comment{grow to $\kappa$ flips within allowance $\beta$}
    \EndFor
\EndFor
\State $G^\star \gets \argmax_{G\in V}\operatorname{AUC}(G)$
\State $\pi^\star \gets$ identity if $G^\star=G_{\mathrm{orig}}$, else
\textsc{Place}$(G^\star)$
\State $F_k \gets$ the set recorded by $G^\star$ at budget $k$, mapped through
$\pi^\star$
\State \Return $(\pi^\star,\{F_k\})$ if the independent check passes
\end{algorithmic}
\label{algorithm}
\end{algorithm} 

\paragraph{Candidate layouts}
A candidate layout specifies which weights occupy $R$.
$G_{\mathrm{orig}}$ leaves $R$ unchanged; other candidates move one reachable row into $R$, where a row is reachable if an admissible permutation can place it in the widest row segment of $R$.
We rate a reachable row by its potential, the sum of its $K_{\max}$ largest positive per-weight scores, and take the top rows under the magnitude and first-order scores.
For the feed-forward output projection, a row is wider than $R$, so all columns are scored and the column permutation is fixed after bit search; other projections score only the columns of the destination segment.

\subsection{Bit Search over Layouts}
Each candidate layout orders its flips in a queue that interleaves the magnitude and first-order rankings at a fixed ratio, one bit per weight. All loss, gradient, and activation measurements, including initialization, are charged to one computational budget $B$. Each candidate keeps a current flip set, the set recorded at each budget, and a cache of marginal gains. A marginal gain is the measured loss increase from adding one flip to the current set; flip sets are
always evaluated jointly, without assuming additive effects.

\paragraph{Successive halving}
A candidate's score, $\operatorname{AUC}$, sums over $k=1,\dots,K_{\max}$ the
largest loss increase it has recorded at any budget up to $k$. Each stage $(c,\kappa,\phi)$ keeps the top $c$ candidates, splits a fraction $\phi$ of the remaining budget equally among them, and grows each to $\kappa$ flips. The first stage ranks candidates by their magnitude potential (the surrogate), which costs no budget; later stages rank by $\operatorname{AUC}$.

\paragraph{Lazy growth}
A candidate first spends a fixed fraction of its allowance measuring the marginal gains of its leading queue entries. It then repeatedly takes the largest positive cached gain, remeasures it if the set has changed since it was cached, and accepts the flip only if the current gain is positive. When no positive gain remains, it measures a few more queue entries. Each accepted set is recorded at the next budget; when growth stops, the current set fills the remaining budgets, so the set recorded at budget $k$ may hold fewer than $k$
flips. After its first accepted flip, a candidate refreshes once if its allowance permits: it recomputes the gradient and activations under the current set and measures a few newly top-ranked flips. Growth uses at most one bit per weight and measures only part of the queue, so it does not guarantee the best flip set.

\subsection{Independent Validation}
\label{sec:method:eval}
The candidate with the largest $\operatorname{AUC}$ determines the layout:
the identity if $G_{\mathrm{orig}}$ wins, otherwise the permutation that
places the selected row, and for the feed-forward output projection its
selected columns, into $R$. The recorded sets are nested, so one permutation
serves all budgets.
Admissibility is decided by an independent evaluator, not by the search. It checks the permutation, the bit budgets, the in-range limit check, and that every
flipped bit lies in $R$. It then requires the clean permuted model to match
the original model within predefined numerical and task tolerances; a layout
that fails this clean gate has all its attacks marked invalid. 

\section{Experimental Evaluation}
\subsection{Experimental Setup}
\label{sec:experimental_setup}
\noindent\textbf{Models.}
We evaluate six LLMs (Llama3.2-1B~\cite{meta2024llama32}, Qwen2.5-1.5B~\cite{yang2024qwen25}, StableLM2-1.6B~\cite{bellagente2024stablelm2}, SmolLM2-1.7B~\cite{allal2025smollm2}, Gemma2-2B~\cite{gemmateam2024gemma2}, and Gemma3-1B~\cite{gemmateam2025gemma3}) and three VLMs (SmolVLM-Instruct~\cite{marafioti2025smolvlm}, Qwen2-VL-2B-Instruct~\cite{wang2024qwen2vl}, and Qwen2.5-VL-3B-Instruct~\cite{bai2025qwen25vl}).
StableLM2-1.6B, SmolLM2-1.7B, and the SmolVLM-Instruct backbone use MHA; Gemma3 uses MQA; the remaining language backbones use GQA.
LLMs are pretrained checkpoints without instruction tuning; VLMs are instruction-tuned.
All language-backbone parameters use symmetric INT8 round-to-nearest quantization, with one FP32 scale per output row or per vector for one-dimensional parameters.
Permutations move codes with their scales, preserving represented values without requantization.

\noindent\textbf{Tasks/Datasets.}
We use WikiText-2~\cite{wikitext2} and MMLU~\cite{mmlu} for LLMs, and OK-VQA~\cite{okvqa} and TextVQA~\cite{textvqa} for VLMs.
Each dataset has an \emph{attack set} supplying losses and gradients for layout and bit search, and a \emph{held-out evaluation set} for clean numerical/task checks, and post-flip evaluation.
The sets share no passage, question, or image.
WikiText-2 uses 128 passages of up to 512 words per set, from training and validation, respectively.
MMLU uses the first five validation questions per subject (285 total) for attack and the remaining 1,246 for evaluation; the five official development examples per subject serve as fixed few-shot demonstrations.
Each VQA dataset uses 128 attack and 256 evaluation image-question pairs from training, with disjoint images.

\noindent\textbf{Metrics.}
We report MMLU accuracy (Acc.), official OK-VQA/TextVQA scores, and WikiText-2 perplexity (PPL).
Relative performance degradation is $\mathrm{RPD}_K=1-(m_K/m_0)^s$, where $m_0$ and $m_K$ are clean and post-flip metrics under the same layout and quantization format, with $s=1$ for Acc./VQA and $s=-1$ for PPL. RPD ranges from $0$ to $1$, increasing with degradation.

\noindent\textbf{Protocol.}
We fix 12 target regions, four each from early, middle, and late decoder layers, for per model and quantization format, 
Each spans 4,096 consecutive stored-weight bytes (4,096 INT8 weights), without assuming physical contiguity or memory-page alignment. Each region receives three runs with different seeds, repeating preparation and bit search under a budget of 300 forward-equivalent units covering forward/backward computation, layout search, and candidate evaluation.
Loss or activation measurements cost one unit; gradients cost three.
Successive-halving stages $(c,\kappa,\phi)$ are $(4,2,0.35)$, $(2,6,0.5)$, and $(1,16,1.0)$.
Candidates comprise the top 24 rows by magnitude and top 8 by
first-order score; flip queues interleave these rankings at $3:1$.
Attack loss is the negative log-likelihood (NLL) of reference tokens
on the attack set, restricted to reference-answer tokens for question
answering.
Each layout remains fixed across $K\in\{0,1,2,4,8,16\}$, with at most
$K$ quantized-weight bit flips; actual counts are reported separately
when search stops early.
At $K=0$, clean agreement with Original must be within 0.01 NLL and
0.02 task score.
Success requires at least doubled PPL or a 50\% relative task-score
reduction.
Comparisons match model, task, quantization, region, $K$, and search budget; metrics are averaged over three runs per region and $K$.
A region favors RegionAlign only if its degradation exceeds both
Original and Random.
All methods share prompts, tokenizers, output limits, and decoding
settings; VQA uses greedy decoding with reported answer truncation.

\noindent\textbf{Baselines.}
We compare RegionAlign with two layouts that share its region-constrained bit search but not its layout choice.
\emph{Original} keeps the model's original layout; \emph{Random} draws one admissible channel permutation per run from the run's seed, outside the adversary's control.
Each runs the same growth on a single candidate, the complete region under its fixed layout, without successive halving.
Random denotes a randomly permuted layout; $K=0$ provides clean controls for all three layouts.

\subsection{Experimental Results}
\label{sec:results}
\subsubsection{\textbf{Clean Performance of RegionAlign Layouts}}
\label{sec:results_clean}
We first evaluate every prepared layout without bit flips ($K=0$) to fix a common starting point for the attack.
Across the 12 target regions and search runs (Table~\ref{tab:clean_checks}), RegionAlign departs from Original by at most 1.12 percentage points on MMLU, 1.41 score points on the VQA tasks, and 0.03 PPL on WikiText-2, and Random is comparable.
All three layouts have near-identical task performance.
This matched starting point lets the following sections attribute post-flip degradation to the selected weight bits rather than to the permutation, since each layout is held fixed while only its in-region bits change.

\begin{table}[t]
\centering
\footnotesize
\caption{Clean performance under three layouts (min--max over 12 target regions
and search runs; accuracy/VQA: 0--100).
RegionAlign closely matches Original.}
\label{tab:clean_checks}

\setlength{\tabcolsep}{2pt}
\renewcommand{\arraystretch}{1.08}
\begin{tabular*}{\columnwidth}{@{\extracolsep{\fill}}lccc@{}}
\toprule
\textbf{LLMs}
& \textbf{Original}
& \textbf{RegionAlign}
& \textbf{Random} \\
\midrule

\multicolumn{4}{@{}l}{\textbf{MMLU: Accuracy (\%)}} \\
\addlinespace[2pt]
Llama3.2-1B
& 30.98 & 30.42--31.78 & 30.42--31.86 \\
Qwen2.5-1.5B
& 59.31 & 58.19--59.95 & 58.35--59.31 \\
Gemma2-2B
& 50.80 & 50.48--51.61 & 50.48--51.52 \\
Gemma3-1B
& 25.60 & 25.12--26.00 & 25.12--26.00 \\
StableLM2-1.6B
& 38.36 & 38.12--38.92 & 38.04--39.33 \\
SmolLM2-1.7B
& 49.92 & 49.84--50.96 & 49.92--51.04 \\

\cmidrule(lr){1-4}
\multicolumn{4}{@{}l}{\textbf{WikiText-2: PPL}} \\
\addlinespace[2pt]
Llama3.2-1B
& 13.55 & 13.55--13.56 & 13.54--13.55 \\
Qwen2.5-1.5B
& 13.16 & 13.16--13.16 & 13.16--13.16 \\
Gemma2-2B
& 11.98 & 11.98--11.98 & 11.98--11.98 \\
Gemma3-1B
& 15.58 & 15.58--15.58 & 15.58--15.58 \\
StableLM2-1.6B
& 11.53 & 11.52--11.53 & 11.52--11.53 \\
SmolLM2-1.7B
& 11.33 & 11.33--11.33 & 11.33--11.36 \\

\midrule
\textbf{VLMs}
& \textbf{Original}
& \textbf{RegionAlign}
& \textbf{Random} \\
\midrule

\multicolumn{4}{@{}l}{\textbf{OK-VQA: VQA score}} \\
\addlinespace[2pt]
SmolVLM-Instruct
& 57.97 & 56.72--58.75 & 56.56--58.75 \\
Qwen2-VL-2B-Instruct
& 59.53 & 58.75--60.31 & 58.36--60.94 \\
Qwen2.5-VL-3B-Instruct
& 60.16 & 59.37--60.55 & 58.98--60.55 \\

\cmidrule(lr){1-4}
\multicolumn{4}{@{}l}{\textbf{TextVQA: VQA score}} \\
\addlinespace[2pt]
SmolVLM-Instruct
& 74.69 & 74.30--75.08 & 74.45--75.04 \\
Qwen2-VL-2B-Instruct
& 84.00 & 83.63--84.80 & 83.63--84.80 \\
Qwen2.5-VL-3B-Instruct
& 87.73 & 86.91--88.01 & 86.56--88.40 \\
\bottomrule
\end{tabular*}
\end{table}

\subsubsection{\textbf{Effect of Target Region and Bit-Flip Budget}}
\label{sec:results_regions}
\begin{figure}[t]
\centering
\includegraphics[width=0.5\textwidth]{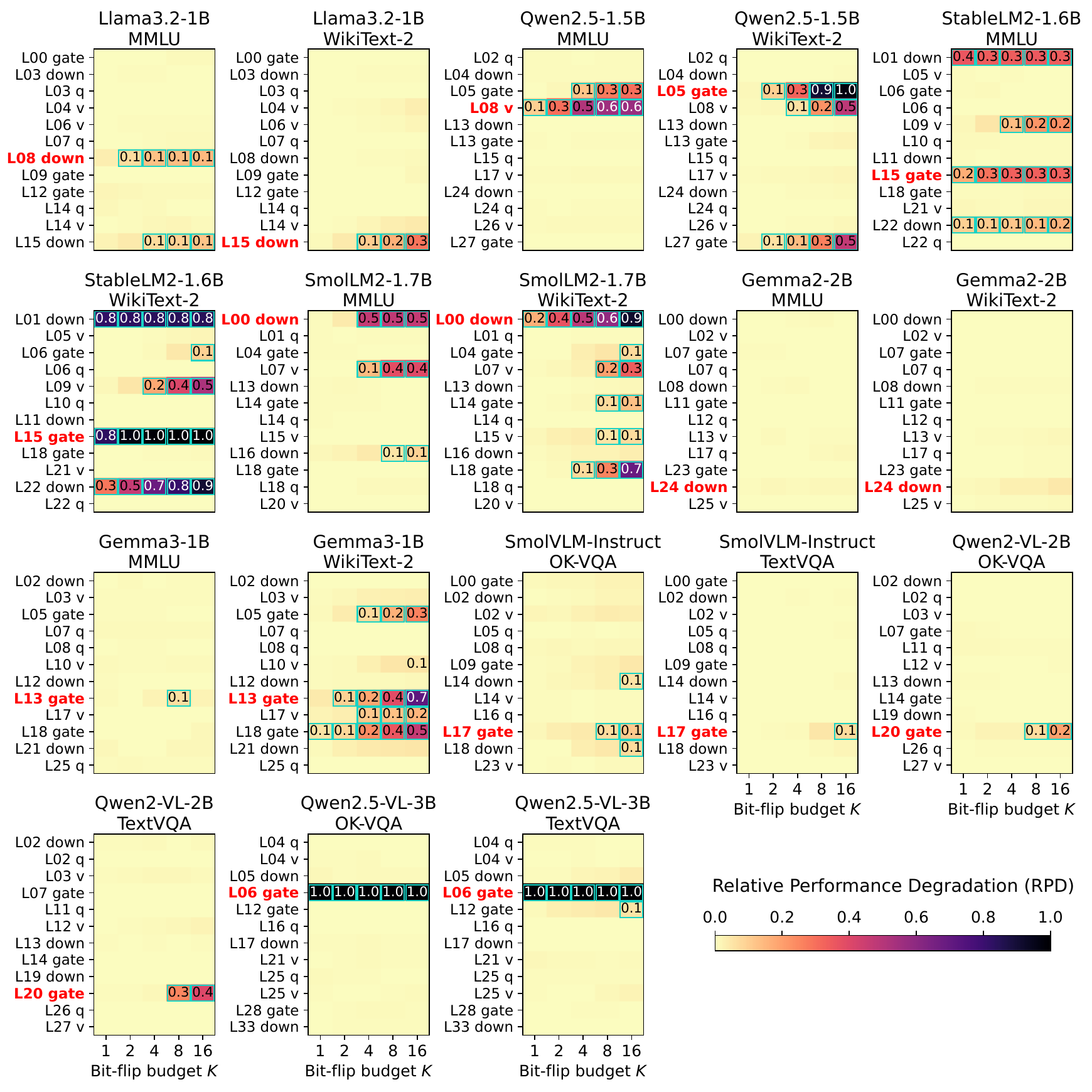}
\caption{RegionAlign RPD across 12 regions and five budgets for all model--task pairs, averaged over three search runs
($0$: no degradation; $1$: full degradation).
Rows show decoder depth and projection type; red marks the region in Fig.~\ref{fig:selected_performance} with the largest advantage over baselines at $K=16$.
Values below $0.05$ are hidden but remain colored (negligible degradation).
Cyan outlines indicate $\mathrm{RPD}\ge0.05$ and higher RPD than in both Original and Random.
Region sensitivity varies across models and budgets.}
\label{fig:region_budget}
\end{figure}

Holding these layouts fixed, we apply the search-selected bit-flip sets to all 12 target regions for every budget $K$.
We call a region \emph{advantaged} when RegionAlign reaches an RPD of at least $0.05$ that is also higher than both Original and Random (the cyan cells in Fig.~\ref{fig:region_budget}).
Per configuration their count ranges from 0 to 6 of 12 and generally grows with $K$ (e.g., from 1 to 6 for SmolLM2-1.7B on WikiText-2), so the attacker-chosen layout helps in a subset of regions that widens with the budget.
The advantage concentrates by projection: at $K=16$ the MLP \texttt{down\_proj} and \texttt{gate\_proj} together with attention \texttt{v\_proj} account for all 39 advantaged regions, whereas \texttt{q\_proj} is never advantaged.
A higher RPD need not cross the success threshold, however: across all evaluations only fifteen regions, spanning early to late decoder layers, let RegionAlign succeed while both baselines fail, and these lie almost entirely on WikiText-2 and the VQA tasks (14 of 15) rather than MMLU.
RegionAlign's benefit thus depends on the target region, the projection, and $K$, motivating a closer look at the regions where it is strongest.

\begin{figure}[t]
\centering
  \includegraphics[width=0.5\textwidth]{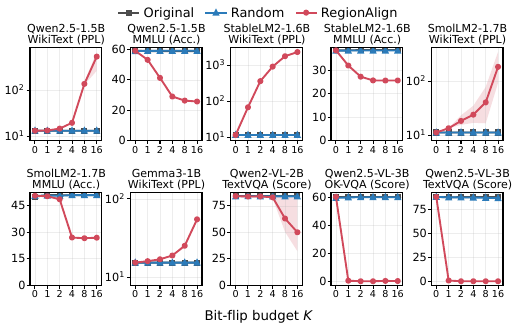}
\caption{Task performance after in-region bit flips. Each configuration shows
the region with the largest RegionAlign advantage over Original and Random
at $K=16$ (red in Fig.~\ref{fig:region_budget}).
Shading shows the range over three search runs; $K=0$ denotes clean performance.
Lower accuracy/VQA scores and higher PPL indicate greater degradation.
RegionAlign causes greater degradation; both baselines remain near clean performance.
}
\label{fig:selected_performance}
\end{figure}

Figure~\ref{fig:selected_performance} shows the region with the largest RegionAlign advantage at $K=16$ per configuration.
Original and Random stay near their clean performance, while RegionAlign degrades more than both baselines in all 18 configurations and crosses the success threshold in seven.
Four bit flips cut Qwen2.5-1.5B's MMLU accuracy to 29.05\% (from 59.31\% clean Original and 59.15\% clean RegionAlign), a single flip raises StableLM2-1.6B's WikiText-2 PPL from 11.53 to 68.29, and a single flip drops Qwen2.5-VL-3B-Instruct below 1 on both VQA tasks.
These drops dwarf the clean differences, so the degradation comes from the in-region bit flips, not the permutation: under the matched region and budget, only the RegionAlign layout exposes such flips.

\begin{figure}
    \centering
    \includegraphics[width=\linewidth]{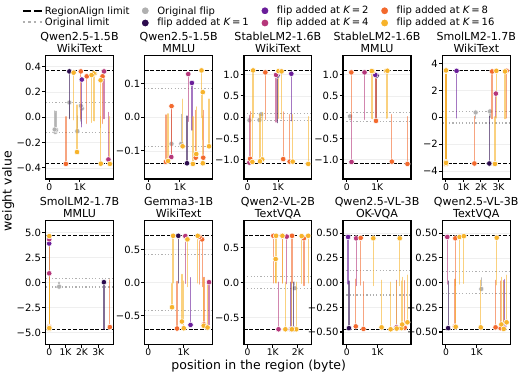}
    \caption{RegionAlign bit-flip mechanism for the model--task pairs in Fig.~\ref{fig:selected_performance}, with the Original layout overlaid.
Each 4\,KiB region holds 4,096 INT8 weights; the x-axis is cropped to the RegionAlign flips.
Each stem runs from a weight's clean value to its post-flip value.
Coloured stems and the black dashed line are RegionAlign (colour marks each flip's first budget $K$; dashed line is its in-range limit); grey stems and the grey dotted line are the Original layout at the same region and budget.
RegionAlign's permutation raises the region's in-range limit and pushes near-zero weights to it by $K{=}1$, while the Original layout barely moves its weights.}
    \label{fig:mechanism}
\end{figure}

Figure~\ref{fig:mechanism} shows how permutation enables damaging attacks with few flips.
Each stem connects a weight's clean and post-flip values, colored by the budget $K$ at which the flip enters, against the row's in-range limit (dashed); Original is overlaid in grey.
Almost all selected flips (149 of 156) target the most-significant bit of near-zero weights, producing large opposite-sign values within range.
These targets exist only after the permutation: under the same region and budget, the Original and Random layouts hold a row with a much smaller in-range limit (e.g., $0.11$ vs.\ $1.10$ for StableLM2-1.6B's layer-15 \texttt{gate\_proj}), so their flips move weights little and barely raise the loss ($\Delta\mathrm{NLL}<10^{-3}$ vs.\ $2.2$); across the selected regions this limit is $1.3$--$12\times$ larger under RegionAlign.
The search uses all $K$ flips except in four MMLU regions (Gemma2-2B, Llama3.2-1B, SmolLM2-1.7B, StableLM2-1.6B), where it stops at $8$--$12$ once no in-region bit increase the loss.
By moving these near-zero, high-leverage weights into the region, RegionAlign makes a few flips far more damaging than either baseline, even though clean behavior is unchanged.

\begin{figure}
    \centering
    \includegraphics[width=0.6\linewidth]{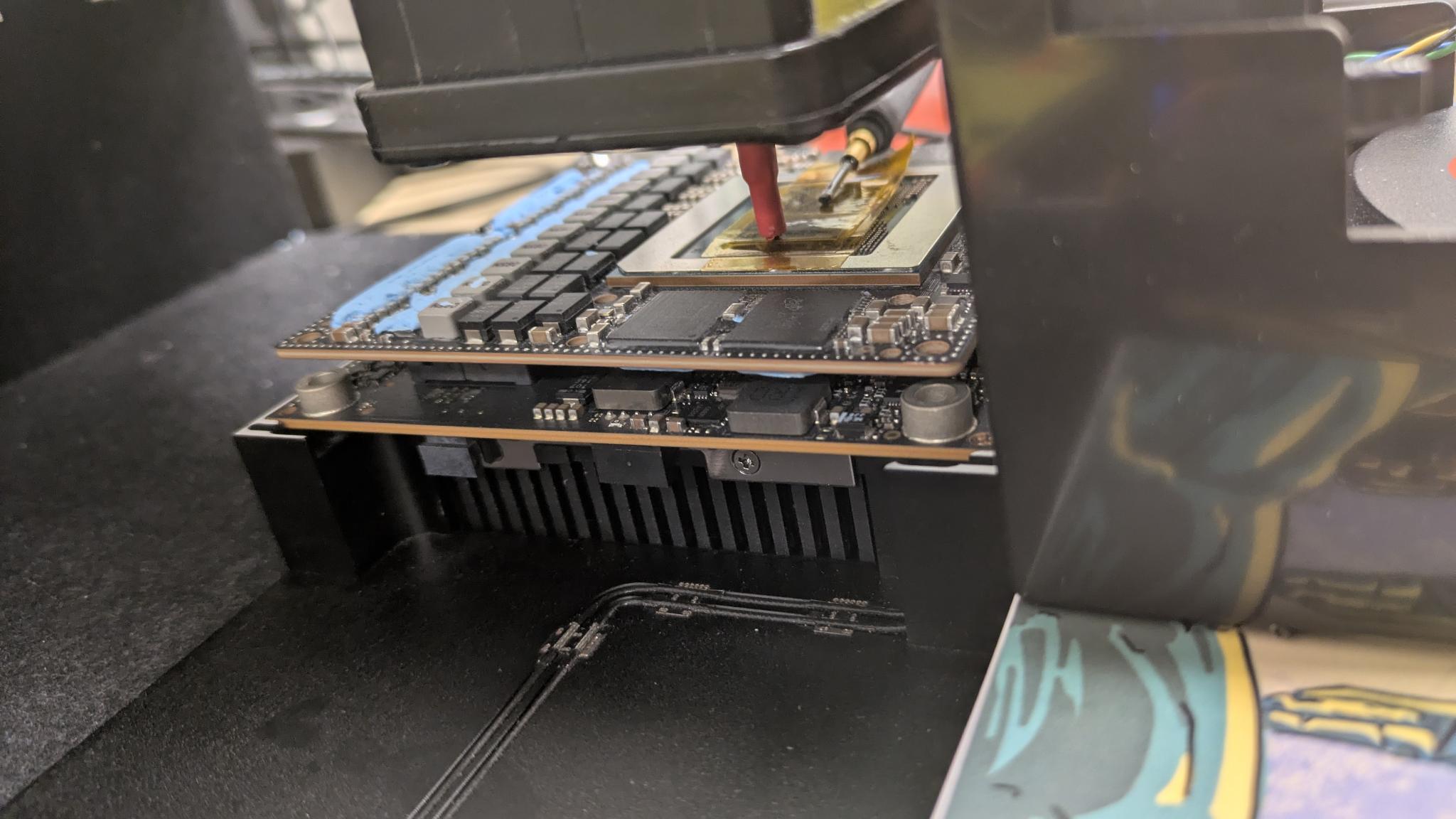}
    \caption{Target device under EMFI: the NVIDIA Jetson AGX Thor SoC die is exposed via removing its thermal enclosure. The die's floorplan is printed and taped to the SoC via Kapton-tape. To inject glitches, we use the Riscure EM-FI transient probe positioned over it, mounted on a motorized XYZ stage with a near-field Langer probe providing a trigger signal. The INT8 weights of Qwen2.5-1.5B reside in the on-board DRAM.}
    \label{fig:hardware}
\end{figure}

\textbf{Hardware Feasibility.}
We deployed Qwen2.5-1.5B (INT8, all parameters quantized) on an NVIDIA Jetson AGX Thor with weights in DRAM, and applied electromagnetic fault injection (EMFI) at the DRAM interface with a Riscure \PH{VC+} transient EMFI probe ($0$--$2$\,V, $10$--$50$\,ns; Fig.~\ref{fig:hardware}). Across $2500$ attempts, $12$ ($\approx0.5\%$) produced confirmed changes in the stored INT8 codes, each a burst of $8$--$112$ codes at $1$--$12$ bits per code (median $\approx4$), with flip directions split almost evenly ($0\!\to\!1$ vs.\ $1\!\to\!0 = 1357{:}1239$) — physical weight faults are real and, as our threat model assumes, direction-agnostic. However, none of the $2500$ attempts hit the registered target (\texttt{layers.8.v\_proj}); steering a fault onto one pre-registered $4$,KiB region required timing sharper than our setup could deliver. We thus report this as a feasibility study: the fault primitive exists, yet un-steered injection lands in a chosen region with probability $\approx0$ ($0/2500$), corroborating our framing that the region is a real attacker constraint and that effectiveness stems from the chosen layout, not from unconstrained physical faults.

\section{Potential Detection and Defenses}
\textbf{Detection.}
RegionAlign nearly preserves clean behavior and leaves weight values and permutation-invariant statistics unchanged.
These properties may hinder output and accuracy monitoring, behavior scanners~\cite{wang2019neuralcleanse}, and other backdoor detectors~\cite{zeng2025clibe,sun2024peftguard}, even with white-box access.
Detection depends on deployment knowledge: a defender with the target region's physical mapping and fault model can reproduce the in-region search to expose vulnerability, potentially at a cost comparable to the attack.
Without this information, assessing physically realizable faults and distinguishing the prepared layout from benign permutations remain difficult.

\textbf{Defenses.}
Potential defenses include weight-integrity checks~\cite{li2021radar,javaheripi2021hashtag},
fault-level checks~\cite{tahmasivand2025lmfix,karamat2025bitflipscope},
software protection~\cite{wang2023aegis}, and decoy
neurons~\cite{liu2023neuropots}.
Validating deployed layouts and dispersing sensitive weights across
regions may also help.
We leave these evaluations to future work.

\section{Conclusion}
We proposed RegionAlign, a supply-chain attack that permutes channels to increase vulnerability to faults in a fixed memory region without retraining or changing weight values or model function.
Software experiments on INT8 LLMs and VLMs show greater degradation than under original or randomly permuted layouts with matched budgets, while preserving clean performance. EMFI induced weight corruption in both directions.
Functional equivalence of the model therefore does not ensure equal fault resilience, motivating validation beyond clean performance.
Future work will explore model preparation that modifies weight values.

\bibliographystyle{IEEEtran}
\bibliography{references,regionalign_terminology_refs}

\end{document}